\documentclass{article}
\usepackage[utf8]{inputenc}

\title{3D Neural Network for Lung Cancer Risk Prediction on CT Volumes}
\author{Daniel Korat}
    
 \date{
    \begin{small}
    The Interdisciplinary Center, Herzliya, Israel\\
    \end{small}
    \begin{small}
    daniel.korat@idc.ac.il
    \end{small}}

\usepackage{natbib}
\usepackage{graphicx}
\usepackage{url}
\usepackage[justification=centering]{caption}

\begin{document}

\maketitle

\section{Introduction}

This project is aimed for the detection of potentially malignant lung nodules and masses. With an estimated 160,000 deaths in 2018, lung cancer is the most common cause of cancer death in the United States \cite{american2014lung}. Lung cancer screening using low-dose computed tomography (CT)
has been shown to reduce mortality by 20–43\% and is now included in US screening guidelines \cite{ardila2019end}. These CT studies can be performed as part of routine screening, for example, studies performed under the National Lung Cancer Screening Trial  (NLST) \cite{nlst}.
Improving the sensitivity and specificity of lung cancer screening is imperative because of the high clinical and financial costs of missed diagnosis, late diagnosis and unnecessary biopsy procedures resulting from false negatives and false positives. The clinical state of the art for diagnosing lung cancer is using the ACR Lung-RADS standard which tries to help a radiologist report in a consistent way and help them decide what is the malignancy risk (and therefore the protocol of treatment).
Despite improved consistency, persistent inter-grader variability and incomplete characterization of comprehensive imaging findings remain as limitations of Lung-RADS. These limitations suggest opportunities for more sophisticated systems to improve performance and inter-reader consistency.
In this project we reproduce a state-of-the-art deep learning algorithm for lung cancer risk prediction. Specifically, our aim is to predict malignancy probability and risk bucket classification from lung CT studies. This would allow for risk categorization of patients being screened and guide the most appropriate surveillance and management.

\subsection{Related Works}

The aim of our project is to reproduce the state-of-the-art \textit{full-volume} model for lung cancer risk prediction, as described in \cite{ardila2019end}. This model is reported to exceed the performance of expert radiologists on this task.

The model’s architecture is an Inflated 3D ConvNet (I3D) \cite{carreira2017quo} which is based on 2D ConvNet \cite{inception-v1} inflation: filters and pooling kernels of very deep image classification ConvNets are expanded into 3D, making it possible to learn seamless spatio-temporal feature extractors from 3D image volumes while leveraging successful ImageNet architecture designs and even their parameters. The reader is referred to the above papers for further details.

\section{Solution}
\subsection{General approach}

Our approach leverages a deep convolutional neural network (CNN) to automate this complex image analysis task. The two main components are the preprocessing module, which segments and centers the lungs in the CT volumes, and the \textit{full-volume} model, a three-dimensional CNN trained to take the entire lung volume and produce a cancer risk prediction score for that patient. Figure \ref{fig:workflow} contains a high-level illustration of the entire workflow.

While \cite{ardila2019end} describe the main implementation details and mention the software tooling used, their code is not made available. We try different methods to obtain the required I3D \cite{carreira2017quo} pre-trained model. Since a compatible version is not available online, we initialize the model ourselves by bootstrapping the filters from the ImageNet-pre-trained 2D Inception-v1 \cite{inception-v1} model into 3D, as outlined in \cite{carreira2017quo}.

We then fine-tune on the preprocessed CT volumes to predict cancer probability within 1 year (binary classification task). These probability scores can then be thresholded \cite{ardila2019end} to assign the appropriate cancer-risk bucket for each patient.

\subsection{Data set}

\begin{figure}
\centering
\includegraphics[scale=0.4]{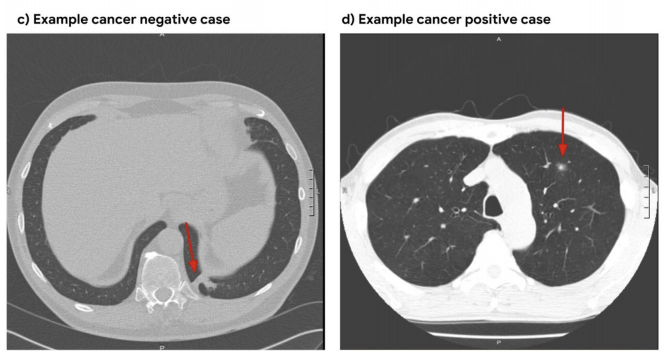}
\caption{Example cancer-negative case with scarring (left) and example cancer-positive case with a nodule.}
\label{fig:example_data}
\end{figure}

Similarly to \cite{ardila2019end}, we use the National Lung Cancer Trial (NLST) dataset \cite{nlst_data}, the largest pubicly available chest LDCT dataset, with 25,000 participants. Every CT image in NLST contains pathology-confirmed cancer evaluations (see Figure \ref{fig:example_data}). Image release is limited to 15,000 participants per project, but due to limited storage and compute time, we use a small subset containing 1,493 volumes (34\% positive).

Obtaining images from the National Cancer Institute (NCI) took around 4 weeks and included submitting a project proposal, a brief review by NCI and completion of a data transfer agreement.
Downloading images is done using the TCIA query tool \cite{tcia}. Finding the correct filter criterion for images was a challenge. We based our filters on \cite{ardila2019end}, but adapted image conditions (e.g. slice thickness and spacing) to the parameters of our distinct preprocessing module (see next section). Identifying the attribute which corresponds to the desired $y$ label (cancer within 1 year or not) required meticulous inspection of the dataset specification sheet. For our full list of query filters, see \cite{github-code}.

\subsection{Design} \label{design}

\begin{figure}
\centering
\includegraphics[scale=0.25]{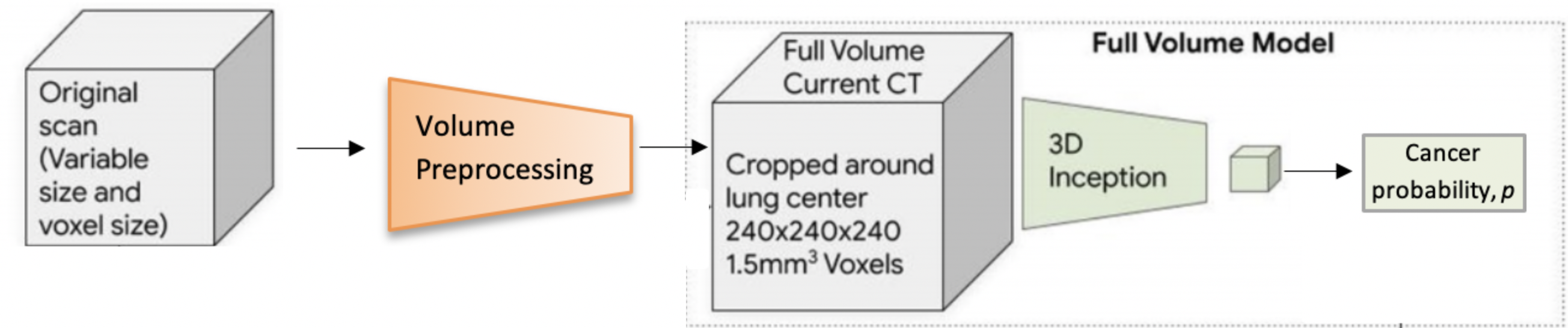}
\caption{High-level illustration of model and data flow.}
\label{fig:workflow}
\end{figure}

\paragraph{Code and Platform.}
We develop our code on Ubuntu 16.04, using Python 3.6. We adopt the I3D model implementation by DeepMind\footnote{\url{https://github.com/deepmind/kinetics-i3d}}. 

The \verb|scikit-learn|\footnote{\url{https://scikit-learn.org}} package is used for image manipulation and metrics such as AUC. See the \verb|requirements.txt| file in \cite{github-code} for the full list of packages used. Training the final model took approximately 4 hours on 2 NVIDIA Titan Xp GPUs.

\paragraph{Preprocessing.}
While \cite{ardila2019end} use a CNN for the preprocessing stage as well, they state that other methods might be preferable. Since the data for training this CNN is not publicly accessible, we employ classic computer vision techniques for this lung segmentation task. When compared to a CNN, our method is explainable, faster to implement, and easily tuned to our model's needs.

Each CT volume in NLST is a directory of DICOM files. Our preprocess module takes such volumes, performs several preprocessing steps, and outputs a fixed-size region cropped around the center of the bounding box computed for the lungs. This 3D RGB matrix can then be fed into the \textit{full-volume} model (see Figure \ref{fig:workflow}). These preprocessing steps are based on this tutorial, and include:\\
\textit{Resampling} to a unified 1.5mm voxel size to reduce unnecessary variance between images.\\
\textit{Lung segmentation}: coarse segmentation used to compute lung center for alignment and reduction of problem space.\\
\textit{Windowing}: clip pixel values using commonly used radiodensity thresholds to focus on lung volume tissue.\\
\textit{RGB normalization}: normalize pixel values to [-1, 1] range which the original Inception network was trained on. 

The last step transforms the data from integer type to float, resulting in substantial growth of image size. Since our storage capacity is limited, we perform this step online (during training) to solve this issue.

\section{Experimental results}

\begin{figure}
\centering
\includegraphics[scale=0.33]{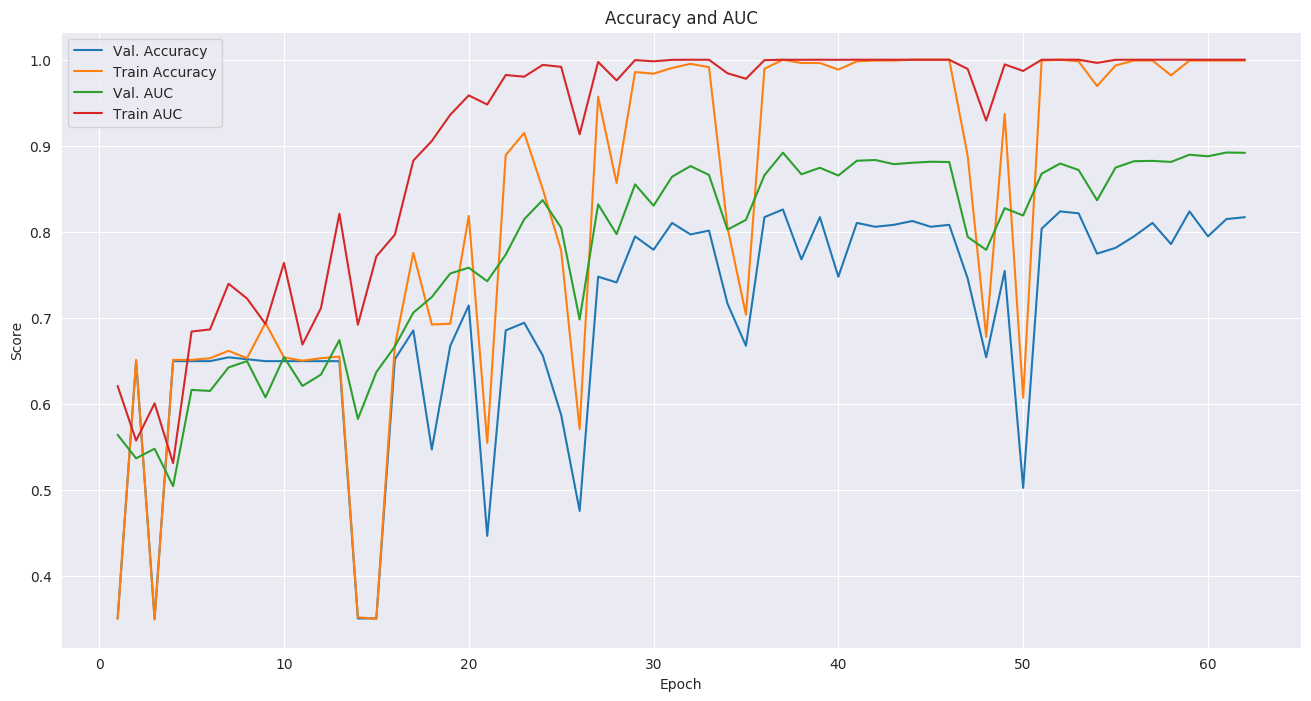}
\caption{AUC and accuracy scores calculated on validation data (green and blue) and training data (red and orange) after each training epoch.}
\label{fig:auc_acc}
\end{figure}

\begin{figure}
\centering
\includegraphics[scale=0.17]{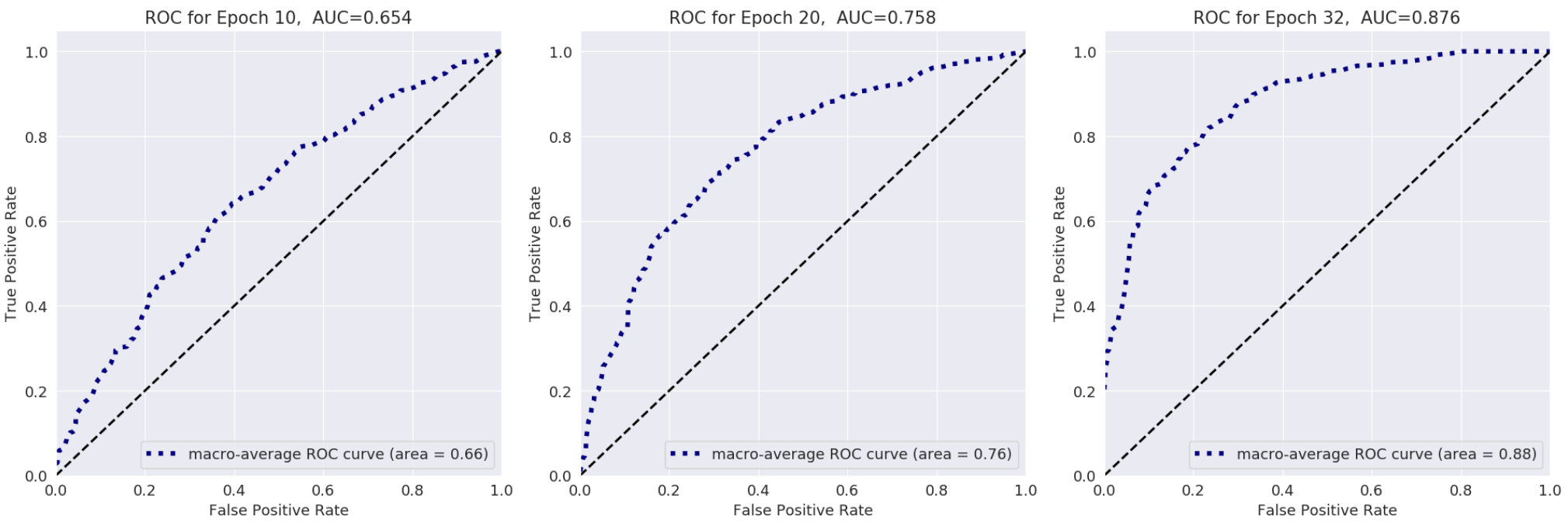}
\caption{Progression of validation ROC curve and AUC score during training.}
\label{fig:roc}
\end{figure}

Following \cite{ardila2019end}, we use Area Under the (ROC) Curve (AUC) as the main evaluation metric. The ROC (Receiver Operating Characteristic) curve illustrates the diagnostic ability of a binary classifier system as its discrimination threshold is varied. The AUC quantifies this plot into a single number between 0 and 1. Figure \ref{fig:roc} shows the ROC curve and AUC score at different epochs during training. We also plot the standard accuracy metric for additional insight. Figure \ref{fig:auc_acc} shows the change in AUC and accuracy as training progresses.

For the training setup, we use a standard 70\% training (1,045 images) 30\% testing (448 images) split. Due to limited GPU memory, we trained in mini-batches of size 2. We used the Adam optimizer with a relatively low learning rate of 5e-5, (due to the small batch size) and halted training before overfitting started around epoch 37. Dropout probability was set to 0.7. Experimenting with different dropout values revealed this is the minimal value which does not cause degraded model accuracy.

We implement the focal loss function used in \cite{ardila2019end}, but did not experience improved results using it, compared to cross-entropy loss which was used instead. The likely reason is that our dataset was more balanced with respect to positive and negative samples. In the preprocessing, we tried different combinations of voxel sizes and lung region sizes. These had to be adapted to fit in the GPU memory, along with the batch size. A batch size of 1 resulted in degraded performance while higher batch sizes and lung regions would not fit in GPU memory. Moreover, resampling to different voxel sizes prolongs the preprocessing duration considerably.

Finally, even though we use a small subset of NLST, we still achieve a state-of-the-art AUC score of 0.892. This is comparable to the AUC for the original \textit{full-volume} model (see the supplementary material of \cite{ardila2019end}), trained on 47,974 volumes (1.34\% positive). Note that in order to obtain a sufficient generalization capability, one would have to train on the full NLST dataset (see instructions in \cite{github-code}).

\section{Discussion}

Our high accuracy results demonstrate that we successfully reproduced the state-of-the-art model. The fact that we achieved the same performance using a training set nearly 50 times smaller is a testament to the effectiveness of the I3D architecture.
When combined with its high consistency and fully automated nature, our approach may enable highly efficient screening procedures and accelerate the adoption of lung cancer screening.

\section{Potential future work}

There are two main interesting possibilities for potential future work:

First, a main point for improvement is accelerating the time for preprocessing and training. While our preprocessing is simple and accurate, it may take up to 7 hours to preprocess 1,000 CT images. Improving computational speed of training can be done, for example, by substituting the underlying Inception-V1 model with one of its improved versions (e.g. \cite{szegedy2017inception}).

Another possibility is to combine our model with valuable patient meta-data (medical history, smoker/non-smoker, demographics, etc.) which exists in NLST. Such meta-data assists radiologists in making more accurate decisions, but is not present in our system.

\section{Code}

Our code and the trained model is available on GitHub\footnote{\url{https://github.com/danielkorat/Lung-Cancer-Risk-Prediction}}, along with instructions for installing, training and predicting using the model. By default, our best fine-tuned model checkpoint is downloaded automatically when executing inference and fine-tuning procedures. This project is also available on   PyPi\footnote{\url{https://pypi.org/project/lungs}} for a one-line installation using \verb| pip install lungs|. Due to usage restrictions, we cannot publish the data. We refer the reader to \cite{nlst_data} for access to the data.

\section{Acknowledgments}

The author thanks the National Cancer Institute for access to NCI's data collected by the National Screening Trial (NLST). The statements contained herein are solely those of the author and do not represent or imply concurrence or endorsement by NCI.

\bibliographystyle{plain}
\bibliography{references}
\end{document}